\documentclass[a4paper,fleqn,usenatbib]{mnras}

\usepackage{newtxtext,newtxmath}

\usepackage[T1]{fontenc}
\usepackage{ae,aecompl}

\usepackage{graphicx}	
\usepackage{amsmath}	
\usepackage{amssymb}	

\title[The possible evolutionary status of {\it BLAP}]{Which evolutionary status does the Blue Large-Amplitude Pulsators stay at?} %
\author[Wu \& Li]{
Tao Wu,$^{1,2,3}$\thanks{E-mail: wutao@ynao.ac.cn}
Yan Li$^{1,2,3,4}$\thanks{E-mail: ly@ynao.ac.cn}
\\
$^{1}${Yunnan Observatories, Chinese Academy of Sciences, 396 Yangfangwang, Guandu District, Kunming, 650216, P. R. China}\\
$^{2}${Key Laboratory for the Structure and Evolution of Celestial Objects, Chinese Academy of Sciences, 396 Yangfangwang, Guandu District, Kunming, 650216, P. R. China}\\
$^{3}${Center for Astronomical Mega-Science, Chinese Academy of Sciences, 20A Datun Road, Chaoyang District, Beijing, 100012, P. R. China}\\
$^{4}${University of Chinese Academy of Sciences, Beijing 100049, China}\\
}

\date{Accepted 2018 May 17. Received 2018 May 7; in original form 2018 March 14}

\pubyear{2018}

\begin{document}
\label{firstpage}
\pagerange{\pageref{firstpage}--\pageref{lastpage}}
\maketitle

\begin{abstract}
   Asteroseismology is a very useful tool for exploring the stellar interiors and evolutionary status and for determining stellar fundamental parameters, such as stellar mass, radius, surface gravity, and the stellar mean density.
   In the present work, we use it to preliminarily analyze the 14 new-type pulsating stars: Blue Large-Amplitude Pulsators (BLAPs) which is observed by OGLE project, to roughly analyze their evolutionary status.
   We adopt the theory of single star evolution and artificially set the mass loss rate of $\dot{M}=-2\times10^{-4}~{\rm M_{\odot}/year}$ and mass loss beginning at the radius of $R=40$ $\rm R_{\odot}$ on red giant branch to generate a series of theoretical models.
   Based on these theoretical models and the corresponding observations, we find that those BLAP stars are more likely to be the core helium burning stars.
   Most of them are in the middle and late phase of the helium burning.
\end{abstract}




\section{Introduction}

{\bf B}lue {\bf L}arge-{\bf A}mplitude {\bf P}ulsator (BLAP) is a new type variable stars observed by the long-term large-scale photometric sky survey the {\bf O}ptical {\bf G}ravitational {\bf L}ensing {\bf E}xperiment (OGLE) \citep[e.g.,][]{Pietrukowicz2013AcA,Pietrukowicz2017NatAs}. The OGLE observation project is established in 1992 and has continuously been providing important discoveries in a variety of fields of modern astrophysics \citep[for more descriptions about the OGLE observation project, please refer to, e.g.,][]{Udalski2003AcA_III,Pietrukowicz2013AcA,Udalski2015AcA_IV,Pietrukowicz2017NatAs}. Since 2010 the project is in its fourth phase (OGLE-IV). Previous phases were: OGLE-I, which is from 1992 to 1995, OGLE-II from 1996 to 2000, and OGLE-III from 2001 to 2009, respectively.

The 14 new type pulsating stars, OGLE-BLAP-001, OGLE-BLAP-002, ..., OGLE-BLAP-014, are listed in Table \ref{tab1_observation}. They have observed in the third (OGLE-III) and the fourth (OGLE-IV) observation phases of the OGLE project \citep[][]{Pietrukowicz2017NatAs}. They are single period pulsation targets with periods ranging from 22 to 39 min. Their relative period changes ($\dot{P}/P$) are the order of $10^{-7}{\rm year}^{-1}$ and the corresponding pulsation amplitudes are of 0.19--0.36 mag in {\it I}-band and of 0.22--0.43 mag in {\it V}-band, respectively \citep[seeing Table 1 of][]{Pietrukowicz2017NatAs}. Phased light curves have a characteristic sawtooth shape \citep[refer to the Figure 1 of ][]{Pietrukowicz2017NatAs}, which is similar to the shape of $\delta$ Scuti stars, classical Cepheids, and the RR Lyrae-type stars pulsating in the fundamental mode. In addition, the pulsating periods and amplitudes are located in the range of $\delta$ Scuti-type stars whose periods ranging from 18 min to 8 h and the amplitude larger than 0.1 mag for the High-amplitude $\delta$ Scuti stars \citep[e.g.,][]{Breger2000ASPC,Niu2013RAA,Fang2016RAA}. Those similar characteristics lead that the prototype object of the whole class, OGLE-BLAP-001, is misidentified as a $\delta$ Scuti-type star and named OGLE-GD-DSCT-0058 \citep[][]{Pietrukowicz2013AcA,Pietrukowicz2017NatAs}. Finally, combining with the spectroscopic observations of targets OGLE-BLAP-001, OGLE-BLAP-009, OGLE-BLAP-011, and OGLE-BLAP-014, \citet[][]{Pietrukowicz2017NatAs} re-identified them as a new type pulsating stars and named them as Blue Large-Amplitude Pulsators (BLAPs). They are located between the $\beta$ Cep and the sdO- and/or sdB-type stars on Hertzsprung-Russell (HR) diagram \citep[refer to the Figure 5 of][]{Pietrukowicz2017NatAs}.

The relative period change $\dot{P}/P$ of pulsating modes is determined by \citet[][]{Pietrukowicz2017NatAs} for 11 of the 14 targets. They are of the order of $10^{-7}~{\rm year}^{-1}$. For the evolutionary status of those stars, \citet[][]{Pietrukowicz2017NatAs} suggested that those stars possibly have lower mass of about 0.3 $\rm M_{\odot}$\ staying at hydrogen-shell-burning, which can be formed in the evolution of a star with the zero-age main-sequence mass $M_{\rm ZAMS}\approx1.0$ $\rm M_{\odot}$\ and losing most of its mass at the red giant branch. Or they have larger mass of about 1.0 $\rm M_{\odot}$\ at core helium burning, which can be formed from the star of $M_{\rm ZAMS}\approx5.0$ $\rm M_{\odot}$.

The stellar evolutionary status can be theoretically determined by the relative period changes, if the relative period changes are dominated by stellar evolution. For instances, the main-sequences (core hydrogen burning), the red-giant (hydrogen-shell-burning), and the horizontal branch stars (core helium burning), even the white dwarf stars (just a cooling process) have different time scales.

In the present work, we will use the theory of the single star evolution to generate a series of models which can reach the region of BLAPs at the HR diagram. From those theoretical models, we will try to analyze and reveal the possible evolutionary status of these new type pulsating targets --- BLAPs.

\section{Observations}
\begin{table*}
\centering
\caption{Summary of the observed properties of 14 known BLAPs, including period ($P$), relative period change ($\dot{P}/P$), effective temperature ($T_{\rm eff}$), surface gravity ($\log g$), and helium--to--hydrogen number ratio ($\log (N_{\rm He}/N_{\rm H})$). These observational parameters are from \citet[][]{Pietrukowicz2017NatAs}, except for the relative period changes of targets  OGLE-BLAP-002, OGLE-BLAP-004, and OGLE-BLAP-011. } \label{tab1_observation}
\begin{tabular}{lccccc} 
\hline\hline
{Variables} & \multicolumn{5}{c}{Values} \\
\hline
{ } & {OGLE-BLAP-001}& {OGLE-BLAP-002} & {OGLE-BLAP-003} & {OGLE-BLAP-004} & {OGLE-BLAP-005} \\
{ } & {OGLE-BLAP-006}& {OGLE-BLAP-007} & {OGLE-BLAP-008} & {OGLE-BLAP-009} & {OGLE-BLAP-010} \\
{ } & {OGLE-BLAP-011}& {OGLE-BLAP-012} & {OGLE-BLAP-013} & {OGLE-BLAP-014}\\
\hline

$P$ [min] & $28.255108\pm0.000105$ & $23.285963\pm0.000002$ & $28.457726\pm0.000002$ &  $22.356690\pm0.000002$ & $27.253491\pm0.000002$ \\
 & $38.015069\pm0.000008$ & $35.181632\pm0.000006$  & $34.481121\pm0.000003$ & $31.935264\pm0.000001$ & $32.132966\pm0.000002$\\
 & $34.874862\pm0.000007$  &  $30.896746\pm0.000002$ & $39.325520\pm0.000004$ & $33.623132\pm0.000002$ \\ \\

$\dot{P}/P$ [$10^{-7} {\rm year}^{-1}$] & $+2.90\pm3.70$ & $-19.23\pm8.05$  & $+0.82\pm0.32$ & $-5.03\pm1.57$  & $+0.63\pm0.26$ \\
& $-3.64\pm0.42$ & $-2.40\pm0.51$ & $+2.11\pm0.27$  & $+1.63\pm0.08$ & $+0.44\pm0.21$ \\
& $+6.77\pm8.87$ & $+0.03\pm0.15$ & $+7.65\pm0.67$  & $+4.82\pm0.39$ \\ \\

$T_{\rm eff}$ [K] & $30800\pm500$  & ... & ... & ... & ... \\
& ... & ... & ... & $31800\pm1400$ & ... \\
& $26200\pm2900$  & ... & ... & $30900\pm2100$  \\ \\

$\log g$ & $4.61\pm0.07$  & ... & ... & ... & ... \\
& ... & ... & ... & $4.40\pm0.18$ & ... \\
& $4.20\pm0.20$  & ... & ... & $4.42\pm0.26$  \\ \\

$\log (N_{\rm He}/N_{\rm H})$ & $-0.55\pm0.05$  & ... & ... & ... & ... \\
& ... & ... & ... & $-0.41\pm0.13$ & ... \\
& $-0.45\pm0.11$  & ... & ... & $-0.54\pm0.16$\\
		\hline
\end{tabular}
\end{table*}

In the work of \citet[][]{Pietrukowicz2017NatAs}, they have determined the periods ($P$) for two OGLE phases, respectively, i.e., $P_{\rm III}$ and $P_{\rm IV}$. Finally, they calculated the relative period changes with the relation of $\dot{P}/P=\frac{\Delta P}{\Delta t}\frac{1}{P_{\rm IV}}=\frac{P_{\rm IV}-P_{\rm III}}{t_{\rm IV}-t_{\rm III}}\frac{1}{P_{\rm IV}}$. For OGLE-BLAP-002, OGLE-BLAP-004, and OGLE-BLAP-011, there are not available light curve in the third observation phase (OGLE-III). Therefore, their relative period changes ($\dot{P}/P$) are unavailable.

Similar to \citet[][]{Pietrukowicz2017NatAs}, we adopt the period of the fourth observation phase ($P_{\rm IV}$) as the final period ($P$) to make analysis. They are listed in Table \ref{tab1_observation}. In addition, we calculate the relative period changes of OGLE-BLAP-002, OGLE-BLAP-004, and OGLE-BLAP-011 from the fourth observation phase with the similar method. We cut the light curves of the fourth observation phase into two parts at the larger gap around the center of the light curves and determine their periods, respectively. Here, the software of Period04 \citep[][]{Lenz2004IAUS..224..786L,Lenz2005CoAst.146...53L,Lenz2014ascl.soft07009L} is used to extract the pulsation frequencies and calculate the corresponding uncertainties from light curves. For OGLE-BLAP-002, we divide the observations into two parts from ${\rm HJD}\simeq2,455,900$ day. The mean moments of the segments are $t_{1}=2,455,556.17641$ and $t_{2}=2,456,277.19568$ day, respectively. Correspondingly, their frequencies are $\nu_{1}=61.8397878\pm0.000071$ and $\nu_{2}=61.8400227\pm0.000068$ c/d, respectively. Finally, we obtain the relative period change of OGLE-BLAP-002 to be of $\dot{P}/P=(-19.23\pm8.05)\times10^{-7} ~{\rm year}^{-1}$. We cut the light curves of OGLE-BLAP-004 at the time of ${\rm HJD}\simeq2,456,600$ day and calculate their mean moments of the segments and frequencies: $t_{1}=2,455,926.19588$, $t_{2}=2,457,154.77045$ day, $\nu_{1}=64.410234\pm0.000016$ and $\nu_{2}=64.4103431\pm0.00003$ c/d, respectively. We finally obtain the relative period change of OGLE-BLAP-004 to be of $\dot{P}/P=(-5.03\pm1.57)\times10^{-7}~{\rm year}^{-1}$. Similarly, for OGLE-BLAP-011, we cut the light curves into two segments at ${\rm HJD}\simeq2,455,900$ day. We calculate their mean moments of the segments $t_{1}=2,455,556.209115$ and $t_{2}=2,456,359.236565$ day and the corresponding frequencies $\nu_{1}=41.29052\pm0.00007$ and $\nu_{2}=41.2904601\pm0.00004$ c/d for the two segments light curves, respectively. Finally, we obtain the relative period change\footnote{In the present work, both of the period change and the relative period change present the relative period change $\dot{P}/P=dP/dt\cdot1/P$.} of $\dot{P}/P=(+6.77\pm8.87)\times10^{-7}~{\rm year}^{-1}$. They are listed in Table {\ref{tab1_observation}}. Compared with the other targets, OGLE-BLAP-002 has larger relative period change in order of magnitude. 

In addition, for OGLE-BLAP-001, OGLE-BLAP-009, OGLE-BLAP-011, and OGLE-BLAP-014, \citet[][]{Pietrukowicz2017NatAs} have made spectroscopic observations and extracted their effective temperature ($T_{\rm eff}$), surface gravity ($\log g$), and helium-to-hydrogen number ratio ($\log (N_{\rm He}/N_{\rm H})$). They are listed in Table \ref{tab1_observation}.

\section{Physical inputs}

In the present work, our theoretical models were computed by the Modules of Experiments in Stellar Astrophysics (MESA), which is developed by \citet{MESA2011}. It can be used to calculate both the stellar evolutionary models and their corresponding oscillation information \citep{MESA2013}. We adopt the package {\small \textbf{``pulse"}} of version {\bf \small ``v6208"} to make our calculations for both stellar evolutions and oscillations \citep[for more detailed descriptions refer to][]{jcd2008,MESA2011,MESA2013}.

Based on the default parameters, we adopt the OPAL opacity table GS98 \citep{gs98} series. We choose the Eddington grey-atmosphere $T-\tau$ relation as the stellar atmosphere model, and treat the convection zone by the standard mixing-length theory (MLT) of \citet{cox1968} with mixing-length parameter $\alpha_{\rm MLT}=2.0$. In addition, the element diffusion, convective overshooting, semi-convection, thermohaline mixing, and the mass-loss due to stellar wind were not included in the theoretical models. However, in order to strip the extra mass from stellar envelope and to generate a series of expected models, we artificially assume the rate of mass ejection ($\dot{M}$) to be a fixed value for a given star.

\section{Modeling and Results}

The spectroscopic observations (helium-to-hydrogen number ratio $\log (N_{\rm He}/N_{\rm H})\sim-0.5$) indicate that the helium-to-hydrogen mass fraction ratio $Y_{\rm surf}/X_{\rm surf}$ is about 1.3 on stellar surface. For a normal star, the surface helium-to-hydrogen mass fraction ratio $Y_{\rm surf}/X_{\rm surf}$ is around $\frac{1}{3}$ at the beginning of stellar formation, such as the Sun. Therefore, those BLAP stars should go through a process of mass loss to strip most mass from stellar surface and leave a similar naked core.

In the present work, the calculations of stellar evolution are started from the pre-main sequence, i.e., the Hayashi line. In order to obtain a suitable model at the expected region of BLAP stars on the HR diagram, similar to the process of generating sdB stars with the way of \citet[][]{Han2002MNRAS.336..449H} and \citet[][]{Xiong2017A&A...599A..54X}, we artificially assume that stars strip their most mass with the mass loss rate of $\dot{M}=-2\times10^{-4}~{\rm M_{\odot}}/{\rm year}$ beginning at the red giant branch with a radius of $R\simeq40~{\rm R_{\odot}}$.

\begin{figure}
   \centering
   \includegraphics[scale=0.47,angle=-90]{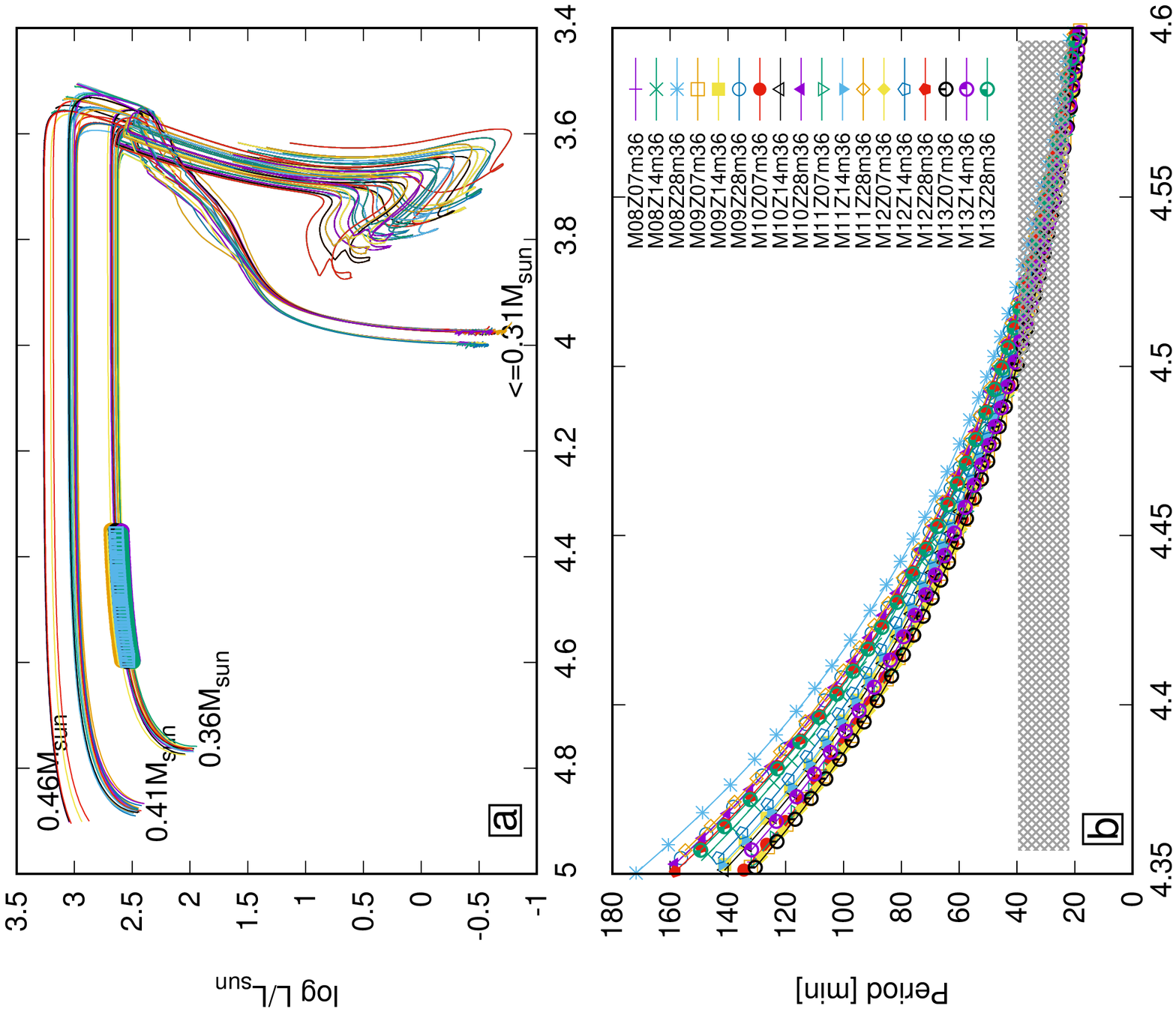}
    \caption{Overview of the model calculations for the low mass case. Upper panel (Panel a) represents the HR diagram of the calculation models whose initial mass $M_{\rm init}\in[0.8,~0.9,~...,~1.3]$, final mass $M_{0}\in[0.26,~0.31,~0.36,~0.41,~0.46]$, and the mass fraction of initial metal abundances $Z_{\rm init}\in[0.007,~0.014,~0.028]$, respectively. The periods and corresponding relative period changes of the pointed models, whose final mass are $0.36~{\rm M_{\odot}}$, are calculated and shown in middle (Panel b) and bottom (Panel c) panels, respectively. In middle panel (Panel b), the shadow area presents the period range of the BLAP, i.e., $22.0\sim39.5$ min. In panel (b), the label ``M08Z07m36" represents initial mass $M_{\rm init}=0.8~{\rm M_{\odot}}$, the mass fraction of initial metal abundance $Z_{\rm init}=0.007$, and final mass $M_{0}=0.36~{\rm M_{\odot}}$.}
    \label{fig:sm}
    \centering
\end{figure}

\begin{figure}
   \centering
   \includegraphics[scale=0.50,angle=-90]{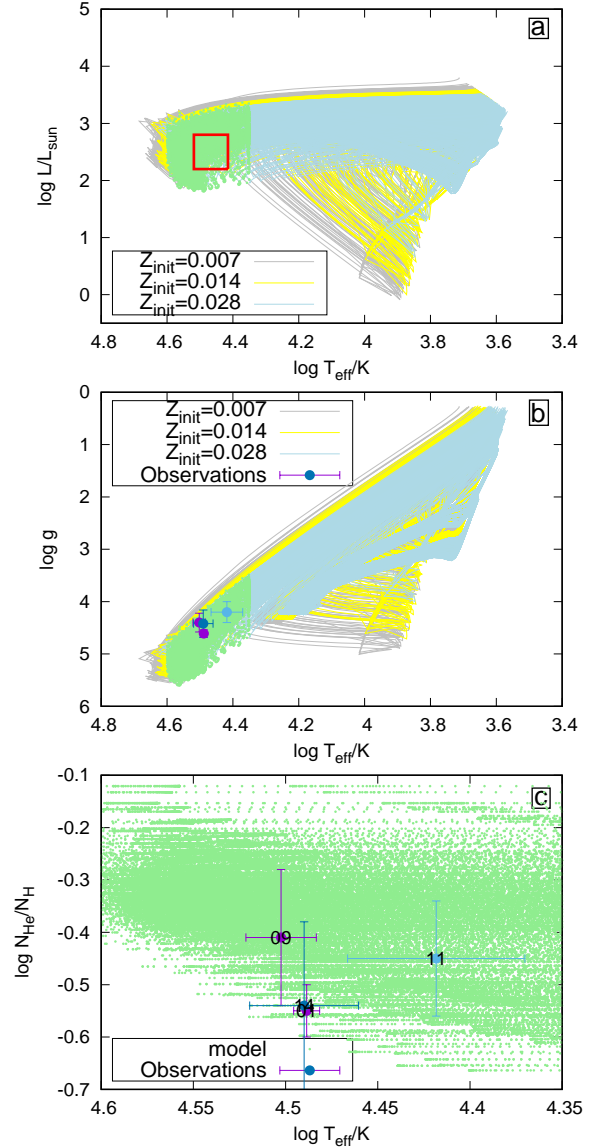}
    \caption{Overview of the model calculations for the larger mass case: Panel (a) -- HR diagram; panel (b) -- $\log g$ vs. $\log T_{\rm eff}$; and  panel (c) -- helium-to-hydrogen number ratio $\log (N_{\rm He}/N_{\rm H})$ vs. $\log T_{\rm eff}$, respectively. In panels (a) and (b), the gray, yellow, and light-blue denote the initial metallicity of $Z_{\rm init}=0.007$, 0.014, and 0.028, respectively. The green points represent the models whose oscillation frequencies are calculated. In panel (a), the red box marks the position of BLAPs at the HR diagram \citep[Figure 5 of][]{Pietrukowicz2017NatAs}. In panels (b) and (c), the points with error-bars are the observations.}
    \label{fig:lm}
    \centering
\end{figure}

\begin{figure*}
    \centering
	\includegraphics[scale=0.51,angle=-90]{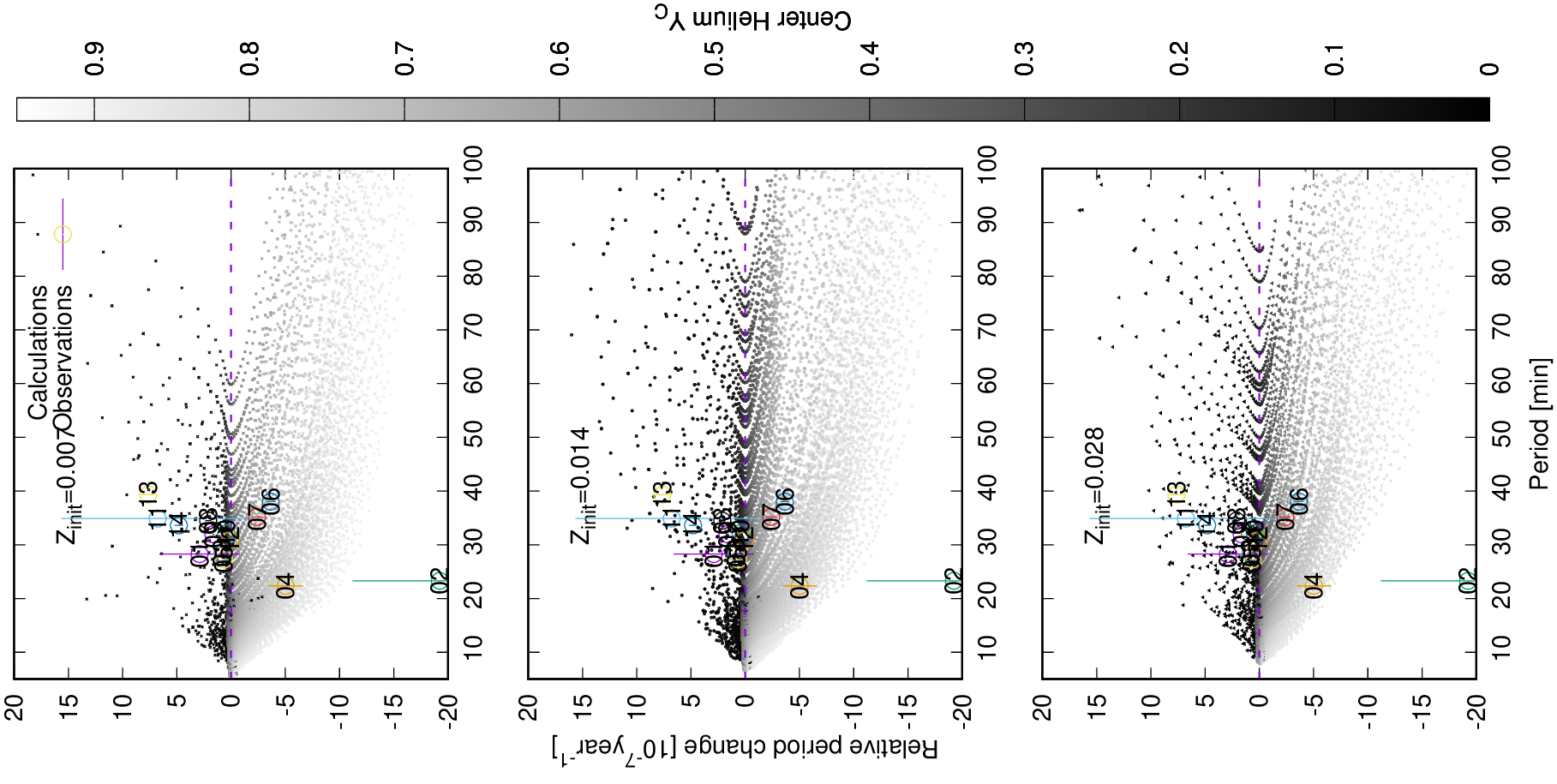}
	\includegraphics[scale=0.51,angle=-90]{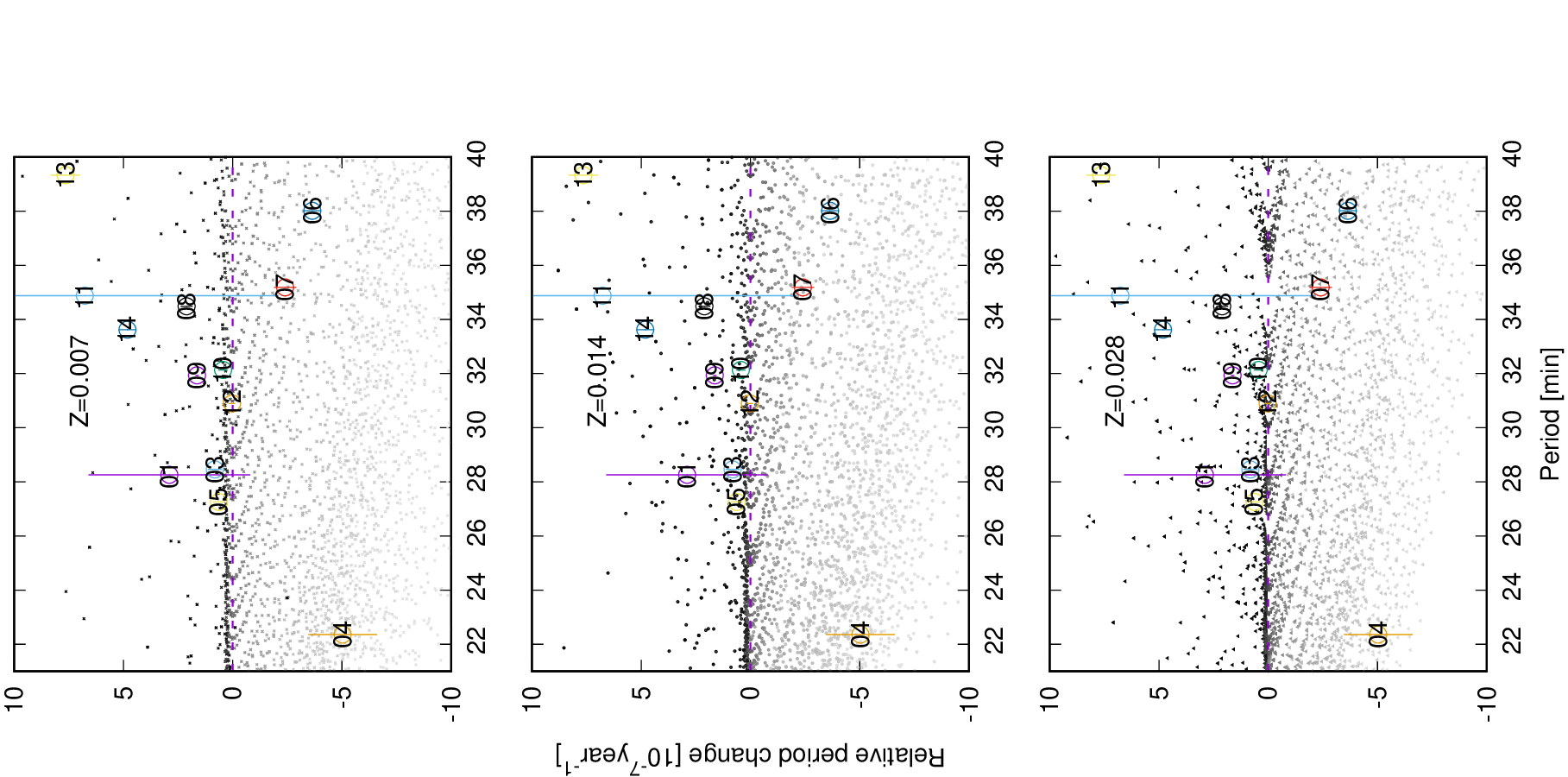}
    \caption{Periods ($P$) vs. relative period changes ($\dot{P}/P$). The gray dots are the model calculations. The color depth of the dots represents the central helium abundance ($Y_{\rm C}$). The purple horizontal dashed lines denote the line of $y=0$. The large color circles present the observations. The corresponding symbolled numbers are the targets ID. The initial metal abundance ($Z_{\rm init}$) gradually increases from upper to bottom panels. They are $Z_{\rm init}=0.007$, 0.014, and 0.028, respectively. The right panels are a zoom of the left panels. }
    \label{fig:dp}
 \centering
\end{figure*}

\begin{figure*}
  \begin{center}
\includegraphics[scale=0.4,angle=-90]{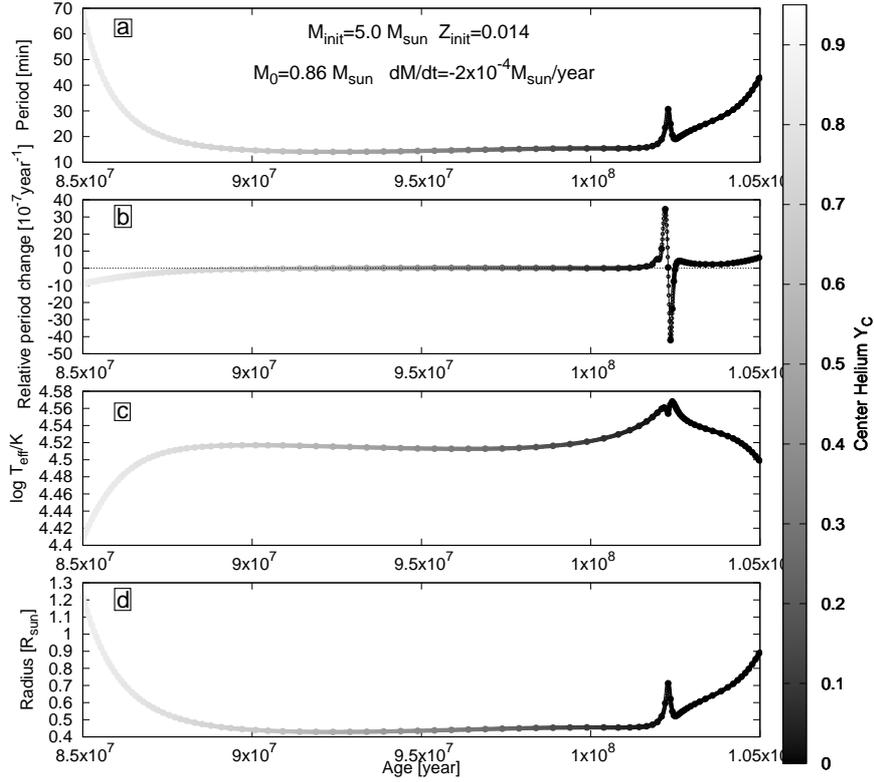}
   \caption{Period ($P$; Panel (a)), relative period change ($\dot{P}/P$; Panel (b)), effective temperature ($\log T_{\rm eff}$; Panel (c)), and stellar radius ($R$; Panel (d)) as a function of stellar age near the BLAP stars at the HR diagram. The color-box represents the mass fraction of helium in stellar center. The large filled point presents the normal calculations and the open point represents the thickened ones by interpolating nine points between two normal calculations, i.e., increasing an order of magnitude for time resolution of model calculations.
  }\label{fig:age.p}
  \end{center}
\end{figure*}

\begin{figure}
  \begin{center}
  \includegraphics[scale=0.56,angle=-90]{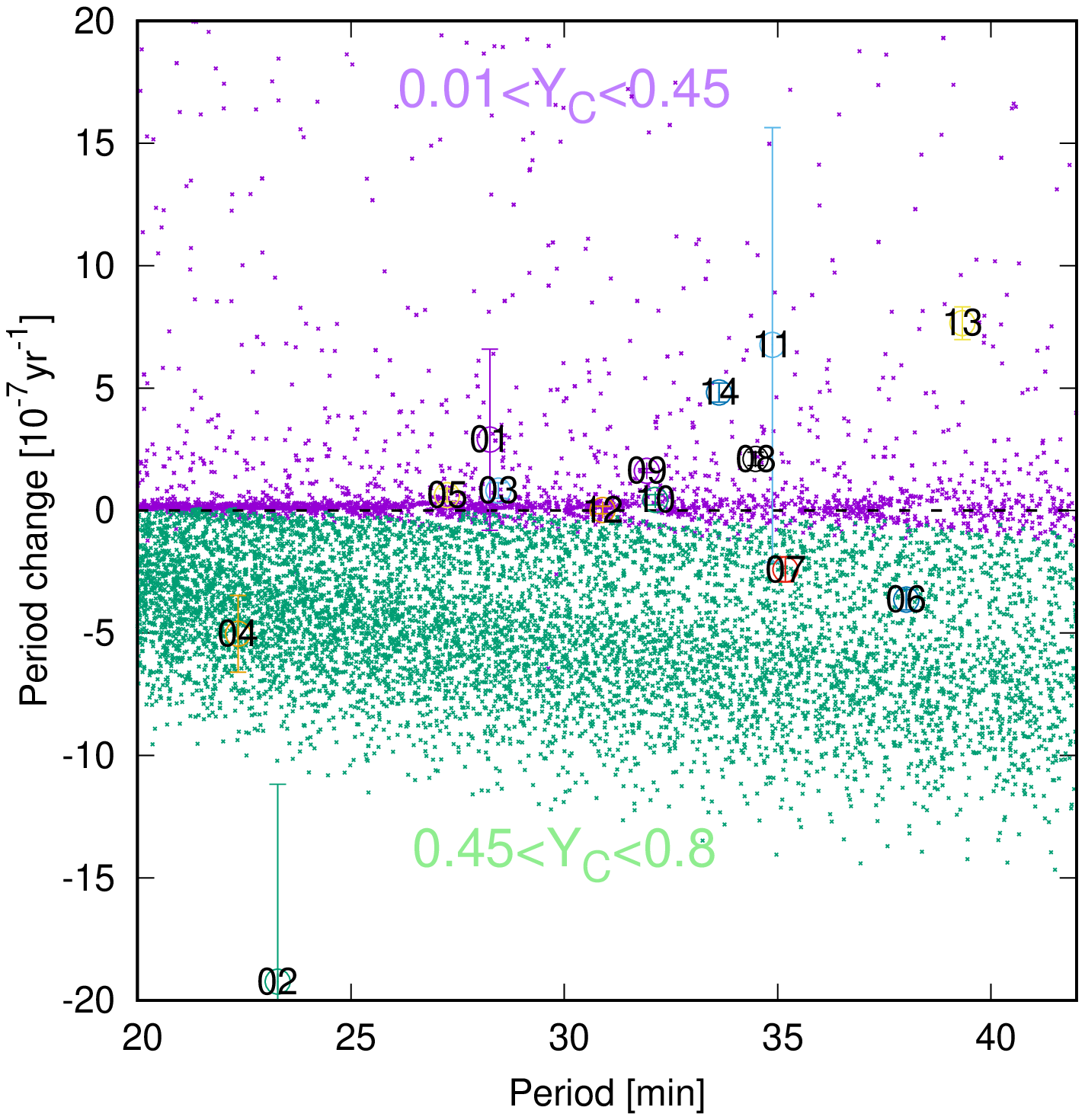}
   \caption{Similar to Figure \ref{fig:dp}, the relative period change ($\dot{P}/P$) vs. period ($P$). The model calculations are divided into two parts. Part I: $0.01<Y_{\rm C}<0.45$ shown with purple points. Part II: $0.45<Y_{\rm C}<0.8$ shown with green points.
  }\label{fig:P.DP.YC}
  \end{center}
\end{figure}

In the work of \citet[][]{Pietrukowicz2017NatAs}, they thought those BLAP stars are low-mass ($\sim0.3$ $\rm M_{\odot}$) stars in the hydrogen-shell-burning phase and/or larger mass ($\sim1.0$ $\rm M_{\odot}$) stars in the helium-core-burning phase. In the present work, we also consider the two corresponding cases: smaller and larger mass. The initial helium abundance $Y_{\rm init}=0.276$, initial metallicity $Z_{\rm init}=0.007$ (metal-poor), $0.014$ (near solar-metal), and $0.028$ (metal-rich), respectively, for both of cases. For the low mass case, the initial mass ($M_{\rm init}$) ranges from 0.8 to 1.3 $\rm M_{\odot}$ with a step of 0.1 $\rm M_{\odot}$ and the final mass ($M_{0}$) from 0.26 to 0.46 $\rm M_{\odot}$ with a step of 0.05 $\rm M_{\odot}$, respectively. They are the hydrogen-shell-burning stars. The calculations are shown in Figure \ref{fig:sm}.

It can be seen from Figure \ref{fig:sm}(a) (HR diagram) that the calculations of those models whose final mass $M_{0}\leqslant0.31~{\rm M_{\odot}}$ are interrupted. They do not arrive at the expected region on the HR diagram. For those model of final mass $M_{0}=0.46~{\rm M_{\odot}}$, some of them evolve to the high effective temperature areas and the other also stay at the lower effective temperature. In addition, these models (final mass $M_{0}=0.41~{\rm M_{\odot}}$) have higher luminosity ($\log L/{\rm L_{\odot}}\sim3$) compared to the observation of BLAPs ($\log L/{\rm L_{\odot}}\sim2.5$). Merely those models whose final mass is $0.36~{\rm M_{\odot}}$ have similar global characteristics with BLAPs on the HR diagram, i.e., similar effective temperature and luminosity. The periods and relative period changes of the corresponding similar models are calculated and shown in Figures \ref{fig:sm}(b) and \ref{fig:sm}(c), respectively. It can be seen from the two figures that the relative period changes ($\dot{P}/P$) are the order of magnitude of $10^{-5}{\rm year}^{-1}$ at the same period ranges with the observations of BLAPs.

For the 14 BLAP stars, their relative period changes are the order of magnitude of $10^{-7}~{\rm or}~10^{-8}~{\rm year}^{-1}$, except for target OGLE-BLAP-002 whose relative period change is at the order of magnitude of $10^{-6}~{\rm year}^{-1}$ ($(-19.23\pm8.05)\times10^{-7}~{\rm year}^{-1}$). Compared with the observations, calculation models have far larger relative period changes on order of magnitude for such low mass case. Therefore, the calculations indicate that the BLAP stars might not belong the hydrogen-shell-burning stars. At least, the above calculations can not meet with observations on the relative period changes. We will calculate and analyze the larger mass case, i.e., stars stay at the core helium burning phase, in the follows.

According to theoretical calculations, we find that those models, whose initial masses are larger than 4.0 $\rm M_{\odot}$\ and the final masses are between 0.5 and 1.2 $\rm M_{\odot}$, might reach the region where those BLAPs are located on the HR diagram, i.e., $\log T_{\rm eff}\simeq4.4-4.5$ and $\log L/{\rm L_{\odot}}\simeq2.2-2.8$ \citep[for more descrptions refer to][]{Pietrukowicz2017NatAs}. In addition, those models are located in the phase of central helium burning. It is worth noting that those models, whose parameters are within the above mentioned ranges but with more hydrogen in its outer envelope, also can not reach this expected region on the HR diagram.

We make calculations for those models whose initial mass ranges from 4.5 to 6.0 $\rm M_{\odot}$ with a step of 0.1 $\rm M_{\odot}$ and the final mass from about 0.66 to 1.21 $\rm M_{\odot}$ with a step of 0.01 $\rm M_{\odot}$, respectively. The composition of elements is the same with the low mass case. The initial helium abundance $Y_{\rm init}=0.276$, initial metallicity $Z_{\rm init}=0.007$ (metal-poor), $0.014$ (near solar-metal), and $0.028$ (metal-rich), respectively. The overview of the calculated models are shown in Figures \ref{fig:lm} and \ref{fig:dp} for larger mass case.

Figure \ref{fig:lm} represents the luminosity $\log L/{\rm L_{\odot}}$ (HR diagram; Panel (a)), surface gravity $\log g$ (Panel (b)), and surface helium-to-hydrogen number ratio $\log (N_{\rm He}/N_{\rm H})$ (Panel (c)) as a function of effective temperature $\log T_{\rm eff}$, respectively. The relation between the relative period changes ($\dot{P}/P$) and periods ($P$) is shown in Figure \ref{fig:dp}.

It can be seen from Figure \ref{fig:lm} that the calculated results are in good agreement with the properties of the observations of targets in global parameters, such as effective temperature ($T_{\rm eff}$), luminosity ($L$), surface gravity ($\log g$), and the surface helium-to-hydrogen number ratio ($\log (N_{\rm He}/N_{\rm H})$). In addition, the model calculations and observations are as well in good agreement with those observations on both of pulsation periods ($P$) and corresponding relative period changes ($\dot{P}/P$) (seeing Figure \ref{fig:dp}). In addition, it can be seen from Figure \ref{fig:dp} that the metal-richer models ($Z_{\rm init}=0.014$ and $0.028$) are slightly better than the metal-poor models ($Z_{\rm init}=0.007$) for marching with observations in relative period change.

The above calculations indicate that, for low mass case (the hydrogen-shell-burning stars), the global fundamental parameters and the pulsation periods of theoretical models are roughly consistent with the observations. But, compared with the observations, they have larger relative period change on order of magnitude. While, all of the available observations are perfectly consistent with the fundamental parameters of the theoretical models for those core helium burning stars. Compared the results between low mass stars with larger mass stars, we find that the BLAP stars are more likely to be the core helium burning phase stars.

As shown in Figure \ref{fig:age.p}, the pulsation period will rapidly decrease with the increase of stellar age at the infancy of the helium-burning, i.e., the period change is less than zero ($\dot{P}/P<0$) and with a relative larger value. And then, the decrease speed of period reduces. The period stays at a stable level and the period change with a smaller value. Finally, the period increases, when the center helium are almost exhausted. During the center helium burning, the variation of stellar radius is fully consistent with period. It can be briefly summarized that stellar radius varies with the center helium burning. Finally, it leads similar variation on pulsation periods. It indicates that the period change contains their inner information. Therefore, the period change can be used to roughly constrain the specific evolutionary status for those BLAP stars.

It can be found from Figures \ref{fig:dp}, \ref{fig:age.p}, and \ref{fig:P.DP.YC} that the relative period changes vary with the center helium abundance $Y_{\rm C}$. The relative period change is less than zero at the beginning of core helium burning. It is larger than zero when center helium ($Y_{\rm C}$) reduces to about 0.45. It can be found from Figures \ref{fig:dp} and \ref{fig:P.DP.YC} that the center helium abundance $Y_{\rm C}$ of the most BLAP stars are less than 0.45. They are in the middle and late phase of the center helium burning.

\section{Conclusions}

BLAPs is a new type pulsating star discovered by the OGLE project \citep[][]{Pietrukowicz2017NatAs}. Thanks to more than 16 years observations, their pulsating periods and the corresponding relative period changes are precisely determined by \citet[][]{Pietrukowicz2017NatAs}. They are identified as the radial fundamental modes. Based on those observations, \citet[][]{Pietrukowicz2017NatAs} suggested that those stars are possible low mass stars ($\sim0.3~{\rm M_{\odot}}$; the hydrogen-shell-burning) and/or larger mass stars ($\sim1.0~{\rm M_{\odot}}$; the core helium burning). In the present work, we generate a series of theoretical models with single evolutionary by MESA \citep[][]{MESA2011,MESA2013} and preliminarily analyze the possible evolutionary status.

Based on the work of \citet{Pietrukowicz2017NatAs}, we make corresponding theoretical model calculations for both of stellar structure and pulsations for both of two possibilities. Finally, we find that these BLAP stars are more likely to be core helium burning phase stars. The order of magnitude of the relative period changes is about $10^{-5}{\rm year}^{-1}$ for the hydrogen-shell-burning stars ($M_{0}\sim0.36~{\rm M_{\odot}}$). It is far larger than the observations which are around $10^{-7}~{\rm or}~10^{-8}{\rm year}^{-1}$. For the core helium burning stars ($M_{0}\sim0.7-1.1~{\rm M_{\odot}}$),  the model calculations are in good agreement with almost all of the available observations, including effective temperature ($\log T_{\rm eff}$), surface gravity ($\log g$), surface helium-to-hydrogen number ratio ($\log (N_{\rm He}/N_{\rm H})$), pulsation period ($P$), and relative period change ($\dot{P}/P$).

We find that the relative period change contain stellar inner information about helium burning from the marching of the relative period change between models and observations. These BLAP stars are in the middle and late phase of the center helium burning. The center helium abundance ($Y_{\rm C}$) is less than 0.45 for most of these BLAP stars. In addition, we find the metal-rich models are better than the metal-poor models for meeting the observations in relative period change.

In the present work, we merely analyze the possible evolutionary status for those BLAP stars. The mass loss ratio are artificially fixed in our theoretical model calculations. There is not enough suitable physical basis to support it.  In the future work, we will consider more physical processes, such as binary evolution, in theoretical models.

\section*{Acknowledgements}

This work is co-sponsored by the NSFC of China (Grant Nos. 11333006, 11503076, 11503079, 11773064, and 11521303), and by Yunnan Applied Basic Research Projects (Grant No. 2017B008). The authors gratefully acknowledge the computing time granted by the Yunnan Observatories, and provided on the facilities at the Yunnan Observatories Supercomputing Platform. The authors also express their sincere thanks to Prof. Zhan-Wen, Han, J.H. Guo, B. Wang, and X.C. Meng, Dr. Q.S. Zhang, J. Su, G.F. Lin, J.Y. Liu, and X.H. Chen, and J.J. Guo for their productive advices. And finally, we thank the referee (Prof. Pawel Pietrukowicz) for constructive comments to improve this paper.












\end{document}